\documentclass{aa}

\usepackage{amssymb}
\usepackage{graphicx}
\usepackage{times}

\begin{document}   

   \thesaurus{06         % A&A Section 6: Form. struct. and evolut. of stars
              (13.07.1)}  % Gamma rays:bursts 
             \title{Archival searches for transient optical emission in 
the error box of the 1991 January 22 gamma-ray burst.}

%   \thanks{}

%   \subtitle{ }

   \author{J. Gorosabel
          \inst{1}
   \and A. J. Castro-Tirado
          \inst{1,2}
          }

   \offprints{J. Gorosabel (jgu@laeff.esa.es)}

   \institute{Laboratorio de Astrof\'{\i}sica Espacial y F\'{\i}sica 
              Fundamental (LAEFF-INTA), P.O. Box 50727,
              E-28080 Madrid, Spain \and Instituto de Astrof\'{\i}sica de
              Andaluc\'{\i}a (IAA-CSIC), P.O. Box 03004, E-18080 Granada}

   \date{Received date; accepted date}

   \titlerunning{Archival searches in the GRB 910122 error box.}

  \authorrunning{Gorosabel \& Castro-Tirado}

   \maketitle

   \begin{abstract}
     We present here the results of a study carried out at the Harvard
     College Observatory Plate Collection. We e\-xamined 3995 plates covering
     the error box of the gamma-ray burst GRB 910122, over a time span of
     90~yr (from 1889 to 1979).  The total exposure time is $\sim$ 0.55~yr.
     No convincing e\-vidence of optical transient emission was found within
     the GRB 910122 IPN error box. However, a possible OT was found {\em
       outside} the GRB 910122 error box.  Optical ground based
     obs\-ervations have revealed a V$\sim$22.3 object consistent with the
     position of the new object.  The colours of the object are ty\-pical
     of a K7/M0 star or a reddened galaxy, which could have caused the OT,
     but the fact that the object is far away from the GRB error box makes
     both events unrelated.  

   \end{abstract}

%
%  14.Sep.'90: Demo-Vs.
%________________________________________________________________

\section{Introduction}
In spite of the exciting progresses carried out recently, the central
engine that powers gamma-ray bursts (GRBs) remains unknown. Following the
detection of an absorption system in the optical spectrum of the optical
transient (OT) related to the gamma-ray burst GRB 970508 (Metzger et al.
1997), it is believed now that most GRBs, if not all, arise at cosmological
distances.  However, the mechanism responsible for the emission is not yet
known, and it could be still possible the existence of recurrent transient
optical emission.

If this is the case, archival searches are extremely important, because
they will allow to search for the possible recurrent optical emission.  The
detection of optical transient emission for GRB 970228 (van Paradijs et al.
1997, Guarnieri et al.  1997) and GRB 970508 (Bond 1997, Djorgovski et al.
1997, Castro-Tirado et al.  1998) confirms this fact.  However, for some
other bursts no optical emission was detected although prompt and deep
follow up observations were carried out (Castro-Tirado et al. 1997, Groot
et al. 1998).

Several OTs have been already found by means of archival searches. However,
the confirmation of optical recurrent emission associated to GRBs is still
unclear. Among them, OT 1901, OT 1928 and OT 1944 (Schaefer 1981, Schaefer
et al. 1984) are located inside small GRB error boxes.  For instance OT
1928 is a prominent object 5.8 mag over the plate limit, within a
~8 arcmin$^2$ error box.  But the reality of these objects has
been under debate (\.{Z}ytkow 1990, Hudec et al.  1994a). Other firm
candidates are the three objects found by Hudec et al. (1990), at the same
position of the sky.  They are well above the plate limit and the structure
is compatible with a star-like object in all three cases.  However, they
are outside the GRB 790325b error box.

Moskalenko's object (OT 1959, Moskalenko et al. 1989) was found inside the
GRB 791101 error box as a 13 mag st\-ellar image (or 6.6 mag if a 1-s flash
is assumed) on two different plates (B and V filters) taken simultaneously.
A double exposure cannot be ruled out because both plates were taken with
cameras on the same mount. Moreover, the low quality and the steepness of
the image profile cast doubts about the reality of the object (Greiner and
Moskalenko 1994).

The OT reported by Greiner et al. (1991), represents one of the best
candidates detected so far.  It was found on three plates taken
simultaneously in 1966 in the GRB 781006b error box. The magnitudes of the
objects are similar to Moska\-lenko's one, at $\sim13.1$ mag (or 6.3 mag if
a 1-s flash is assumed). Although some preliminary explanations were given
(asteroids, satellite glints, head-on meteors, aircraft air lights or a
dwarf nova), spectroscopic observations provided evidence that the OT was
due to a large amplitude flaring dMe star, probably unrelated to the GRB
(Greiner and Motch 1995).

The only OT found near-simultaneous to a GRB event, is the one reported by
Borovi\u{c}ka et al. (1992). They found a star-like object at the edge of
the GRB 790929 error box on a plate taken only 7.1 h after the high energy
event.  Also, two additional objects were found at the same position on
another plates.  The objects are consistent with the star HDE 249119, but
are relatively faint (at the utmost 1 mag above the plate limit), so a
plate fault or a grain cumulation cannot be completely excluded.  Further
optical studies detected additional optical eruptions of HDE 249119,
reporting ten optical brightenings with amplitudes greater than 0.5 mag
(\u{S}t\u{e}p\'an and Hudec 1996).

Besides the above-mentioned OT candidates, Hudec et al. (1994b) reported
the presence of a very bright OT (5 mag above the plate limit) inside GRB
910219 WATCH/{\it Gra\-nat} error box (Sazonov et al. 1998) coincident with
the position of a confirmed quasar with redshift z=1.78 (Vrba et al. 1994).
However the position of the object is not consistent with the small error
box derived by the third Interplanetary Network (hereafter IPN, Hurley
1997).

More recently Hudec et al. (1997) reported two OTs that may be connected to
GRB 910522 and GRB 920406. In the former case the OT seems to be related to
a dMe flare star whereas a quiet counterpart with colours consistent with a
quasar/AGN was found in the second one.

Other OT candidates have been found (Scholz 1984, Greiner and Flohrer 1985,
Flohrer et al. 1986, Greiner et al.  1987, Hudec et al. 1987, Hudec et al.
1988) but none of them can be conclusively proven as a real OT. Therefore,
despite strong efforts carried out during the last two decades there is no
a convincing prove that OTs found in plate archives and GRBs are related to
the same physical phenomenon. Comprehensive reviews on OTs found so far can
be seen in Hudec et al.  (1993a, 1993b, 1994c, 1995).

The most recent archival search has been performed by Gorosabel and
Castro-Tirado~~(1998a).~~They have examined 8004 plates at the Harvard
College Observatory Plate Collection searching for optical transient
emission from the gamma-ray burst GRB 970228.  This has been the first
archival search carried out so far for a gamma-ray burst with known
transient optical emission.  The total exposure time amounted to $\sim 1.1$
yr. No convincing optical activity has been found above $12.5$ mag at the
expected position of the GRB 970228 optical counterpart. Table 1 displays a
summary of the most exhaustive archival searches performed to date.

\begin{table}[t]
\begin{center}
\caption{The most exhaustive GRB archival searches carried
  out to date.}
\begin{tabular}{lccc}
  \hline
Plate      & Number  &    Total         & Reference  \\
collection & of      &   monitoring     &            \\
           & boxes   &    time ({\scriptsize yr})     &            \\
   \hline                
   Ond\u{r}ejov          & 22& $\sim$12 & Hudec et al.    \\
   Sonneberg             &   &          & (1987, 1988, 1991, 1994b)\\
   Bamberg               &   &          &        \\
                         &   &          &        \\
   Harvard               & 16&      4.25& Schaefer et al.\\
                         &   &          & (1984, 1990)   \\ 
                         &   &          &        \\
   Sonneberg             & 15&      2.6 & Greiner et al.\\
                         &   &          & (1987, 1990b, 1990c) \\
                         &   &          &        \\
   Harvard               &  2&      1.65& Gorosabel \& Castro-Tirado.\\
   ROE                   &   &          & (This paper and 1998a)\\
                         &   &          &        \\
   Odessa                & 40&      1.26& Moskalenko et al.\\
                         &   &          & (1989, 1992) \\
\hline   
\end{tabular} 
\end{center}
\end{table}

\begin{table}[t]
\begin{center}
\caption{The 7 HCO Plates examined for GRB 910122.}
\begin{tabular}{lccccl}
\hline
{\small Plate}&{\small Limiting}&{\small Number}&{\small Total}&{\small Number}&{\small Faults} \\
{\small series}&{\small magnitude}&{\small of} &{\small exposure}&{\small of plate }&{\small rate}\\
& &{\small plates}&{time \scriptsize (hr)}&{\small faults}& {\scriptsize (mm$^{-2}$)}\\
\hline
{\small A    }&{\small 16.5}&{\small 16  }&{\small 23  }&{\small 1 }&{\small $5.0~10^{-5}$} \\
{\small AM   }&{\small 14.1}&{\small 3001}&{\small 4078}&{\small 94}&{\small $1.2~10^{-3}$} \\
{\small B    }&{\small 15.2}&{\small 277 }&{\small 78  }&{\small  8}&{\small $1.6~10^{-4}$} \\
{\small Damon}&{\small 15.2}&{\small 254 }&{\small 285 }&{\small  5}&{\small $1.1~10^{-4}$} \\
{\small MF   }&{\small 16.4}&{\small 135 }&{\small 82  }&{\small  6}&{\small $2.1~10^{-4}$} \\
{\small RB   }&{\small 14.8}&{\small 312 }&{\small 282 }&{\small 11}&{\small $9.8~10^{-4}$} \\
\hline
\end{tabular} 
\end{center}
\end{table}

GRB 910122 is one of the GRBs with a very small error box (see Fig. 1).
The ${\it X}$-ray fluence in the 8-20 keV range, as seen by WATCH
(Castro-Tirado 1994) was $2.1\times10^{-6}$ erg cm$^{-2}$, and the
${\gamma}$-ray fluence (E $ > $ 100 keV) was $6.77\times10^{-5}$ erg
cm$^{-2}$ (Therekov et al.  1994).  The detection of GRB 910122 by the IPN
(that included {\it Ulysses}, SIGMA/{\it Granat} and {\it PVO} spacecraft)
allowed the intersection of three annuli which determined a tiny error box
of $\sim 20 $~arcmin$^{2}$ (Hurley et al. 1993). The position of the error
box was afterwards improved, being shifted from the previous IPN position
$~4^{\prime}$ to the west (Hurley 1997). The new IPN error box is
consistent with that provided by WATCH/{\it Granat} and is the basis of our
study.

In this paper we report the results of a study carried out at several plate
archives, looking for optical transient emission from GRB 910122.
Preliminary results have been published elsewhere (Guziy et al. 1997).

\section{Observations and results}
\subsection{No optical transient found within the IPN error box}

Both the small size of the GRB error box and the low surface density of
stars ($b^{\rm II}=-30^{\circ}$) made the search at the Harvard College
Observatory Plate Collection (HCO) very suitable.  A total of 3995 plates
were examined (see Table 2). A region $\sim 100$ times larger than the
$3\sigma$ error box around the GRB position was visually inspected, that
was used to estimate the star-like plate faults rate (see Table~2).  The
total exposure time for GRB 910122 amounts to $\sim$~0.55~yr.

\begin{table}[t]
\begin{center}
\caption{The 7 HCO plates containing objects selected for a deep study.}
\begin{tabular}{lcr}
  \hline 
 Plate ID &  Plate       & exposure \\ 
 number   &  Series      & (\scriptsize minutes)\\ 
  \hline 
 2027     & Damon Blue   & 30 \\ 
 299      & Damon Yellow &120 \\ 
 24633    & AM           & 90 \\ 
 24098    & AM           & 60 \\ 
 24589    & AM           & 90 \\ 
 108      & AM           & 63 \\ 
 45398    & B            & 10 \\ 
\hline
\end{tabular}
\end{center}
\end{table}

In this study 125 suspicious objects were found, being 118 easily discarded
using amplification lenses. The remaining 7 objects (see Table 3) were
checked using the reflected-transmi\-tted light microscope. None of the 7
objects are located {\em within} the 910122 IPN error box.  Most of them
are located far away from the error box. Only the closest object to the GRB
error box (on plate AM 24589) was considered for further study.

Additional plates taken at the UK Schmidt Telescope (UKST) in Australia
were blinked at the Royal Observatory of Edinburgh (ROE). No object varying
by more than 0.2 mag was found near or inside the IPN error box above the
plate limit, 20-21 mag.  Table 4 displays the plates examined at ROE.

\begin{table}[t]
\begin{center}
\caption{The 18 UKST plates examined at ROE.}
  \begin{tabular}{lcr}
\hline
   Plate ID & Date & Exposure  \\
   number &      & time (s)   \\
 \hline                        
 UB   53  & 730727 &   200  \\
  J  897  & 740908 &   400  \\
  J  916  & 740916 &   450  \\
  J 1751  & 750807 &   700  \\
  J 1758  & 750808 &   700  \\
  J 1791  & 750829 &   700  \\
 HA 2560  & 760829 &  1300  \\
 HA 2585  & 760912 &  2400  \\
 BJ 4251  & 780506 &   150  \\
  J 7019S & 810511 &   600  \\
  J 7024  & 810512 &   700  \\
  J 7192  & 810917 &   100  \\
 UJ10444P & 850911 &   250  \\
  B11967  & 870622 &   150  \\
  U11968  & 870622 &   600  \\
 OR14496  & 910831 &   650  \\
 OR15122  & 920821 &   600  \\
  I15763  & 930912 &  1200  \\
 \hline      
   \end{tabular} 
\end{center}
\end{table}

\subsection{An object found on plate AM 24589 close to the error box} 

The plate AM 24589 is a 90-min blue exposure taken in 1945 May 18 where the
object appears as a $13.5 \pm 0.5$ mag star, or $\sim 0.6$ mag above the
plate limit. The object is lightly more compact than the stars on the
plate.  The plates taken close in time to AM 24589 were AM 24582 (exposed
the night before) and AM 24591 (exposed 3.05 hr after). None of them showed
something at the object position (Fig.  1). Like AM 24589, both plates were
exposed for 90 minutes, thus the limiting magnitude is the same (B $\sim
14.1$).

 \begin{figure}[h]
  \begin{center}
    \resizebox{\hsize}{!}{\includegraphics{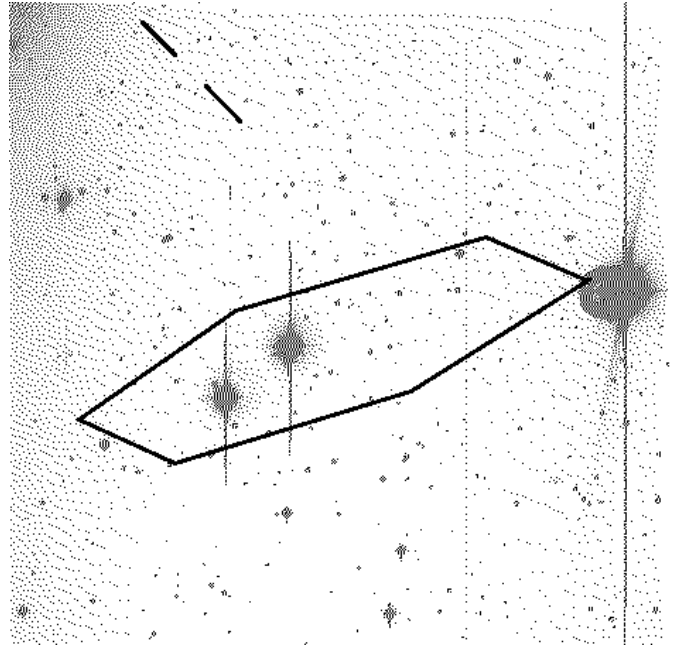}}
    \caption{A B-band image obtained with the 1.54-m Danish Telescope at La
      Silla. The improved error box of GRB 910122, provided by the third
      interplanetary network (Hurley 1997), and the location of the
      possible optical transient are shown.  The field of view is
      $12^{\prime}.6 \times 12^{\prime}.6$.  North is at the top and east
      to the left.}
  \end{center}
 \end{figure}

\subsection{Optical observations of GRB 910122 error box}
Optical observations were performed in 1996-1997 during three observing
runs on the Danish 1.54-m Telescope at ESO La Silla Observatory. The DFOSC
instrument provided a field of view of $13^{\prime}.3 \times
13^{\prime}$.3. The CCD was a backside illuminated Loral/Lessler chip with
2052$\times$2052 pixels.

All observations were made using UBVR (Bessel) and I (Gunn) filters.
Photometric calibrations were made using the Landolt field Mark A (Landolt
1992), and the standard stars HD 84971 and HD 121968 at different air
masses and filters. Limiting magnitudes for the co-added images are I=22,
R=23.5, V=24, B=24.5, U=22. Table 5 displays the observing log.

The IPN error box contains $\sim$ 150 objects brighter than V$\sim$22 mag.
There are two bright stars (V$\sim$10) within the error box. None of the
$\sim$ 150 objects contained in the error box seem to be elongated,
although in the surrounding area, close to the edge of the error box, there
are four galaxies with V$\leq$19. At first sight there are no objects
within the IPN error box with anomalous colours. A detailed study about the
content of the error box will be published elsewhere (Gorosabel,
Castro-Tirado and Lund 1998).

\begin{figure*}[t]
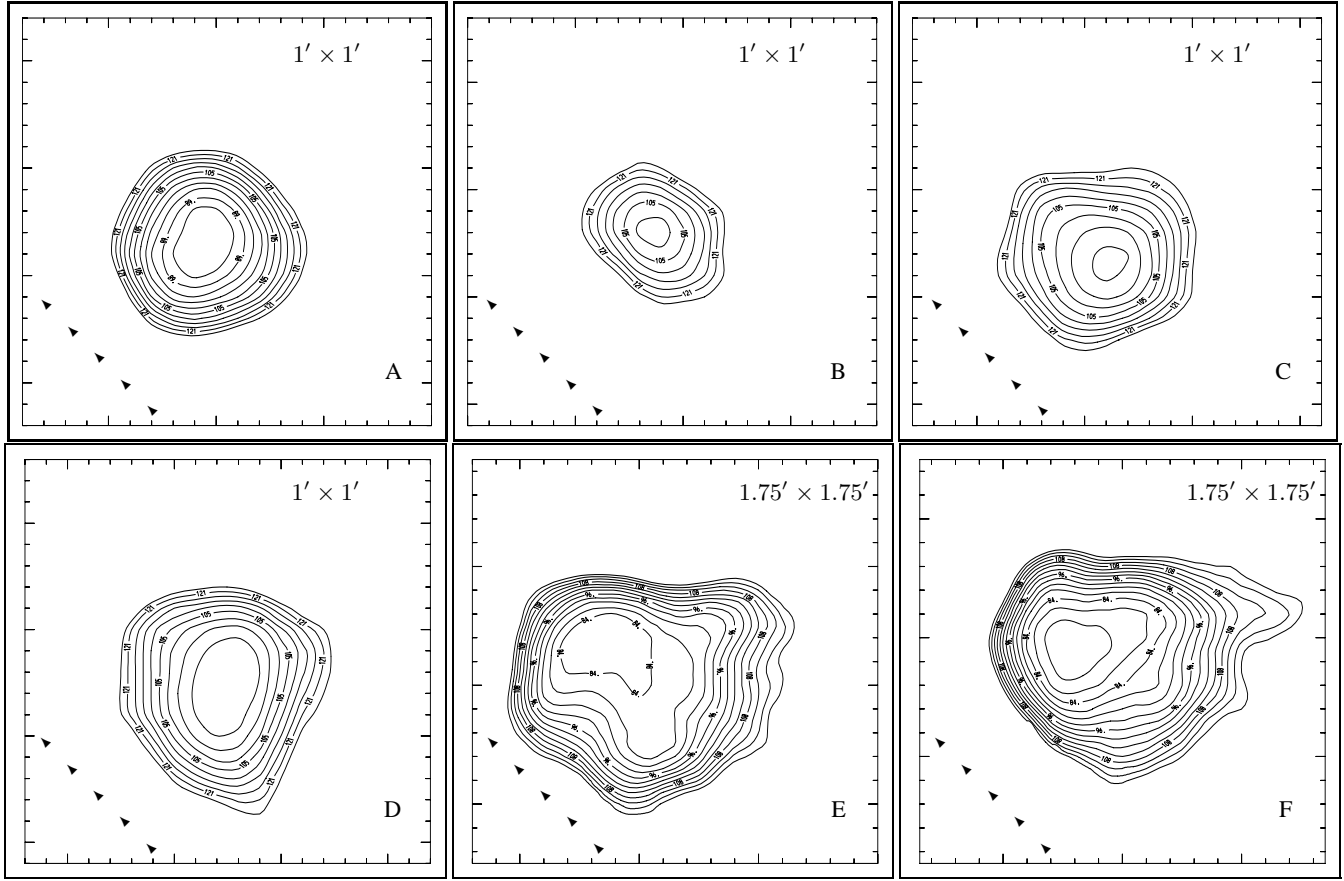

  \centering 

\fbox{\resizebox{5.6cm}{5.6cm}{\includegraphics{7418.f2}}
\put(-20,20){A}
\put(-55,140){$1^{\prime}\times1^{\prime}$}
\put(-100,0){\vector(-2,2){50}}}
%\fbox{\resizebox{5.6cm}{5.6cm}{\includegraphics{new_23_smooth.eps}}} 
\fbox{\resizebox{5.6cm}{5.6cm}{\includegraphics{7418.f3}}
\put(-20,20){B}
\put(-55,140){$1^{\prime}\times1^{\prime}$}
\put(-100,0){\vector(-2,2){50}}} 
\fbox{\resizebox{5.6cm}{5.6cm}{\includegraphics{7418.f4}}
\put(-20,20){C}
\put(-55,140){$1^{\prime}\times1^{\prime}$}
\put(-100,0){\vector(-2,2){50}}}
%\fbox{ \resizebox{5.6cm}{5.6cm}{\includegraphics{29_smooth.eps}}}
\fbox{ \resizebox{5.55cm}{5.55cm}{\includegraphics{7418.f5}}
\put(-20,20){D}
\put(-55,140){$1^{\prime}\times1^{\prime}$}
\put(-100,0){\vector(-2,2){50}}}
\fbox{ \resizebox{5.55cm}{5.55cm}{\includegraphics{7418.f6}}
\put(-20,20){E}
\put(-55,140){$1.75^{\prime}\times1.75^{\prime}$}
\put(-100,0){\vector(-2,2){50}}}
\fbox{ \resizebox{5.55cm}{5.55cm}{\includegraphics{7418.f7}}
\put(-20,20){F}
\put(-55,140){$1.75^{\prime}\times1.75^{\prime}$}
\put(-100,0){\vector(-2,2){50}}}

\caption{Image ``A'' shows the equidensity profiles as detected by the 
  transmitted light microscope for the object found on the plate AM 24589.
  The remaining images show equidensity diagrams for nearby stars on the
  same plate. Images ``B'', ``C'' and ``D'' show the profiles for faint
  stars (B$\sim$13-14), whereas the contours of two bright stars
  (B$\sim$9-10) are shown in ``E'' and ``F''.  The contours are labeled in
  arbitrary units which are directly related to the intensity of the light
  measured by the CCD at the microscope.  The direction towards the plate
  centre is indicated by the arrows. Fields of view are indicated at the
  upper right corners. As it is shown, the object found on the plate AM
  24589 shows similar profiles to those measured in stars with similar
  magnitudes to it.  North is at the top and east to the left. }

\end{figure*}

\begin{figure*}[t]
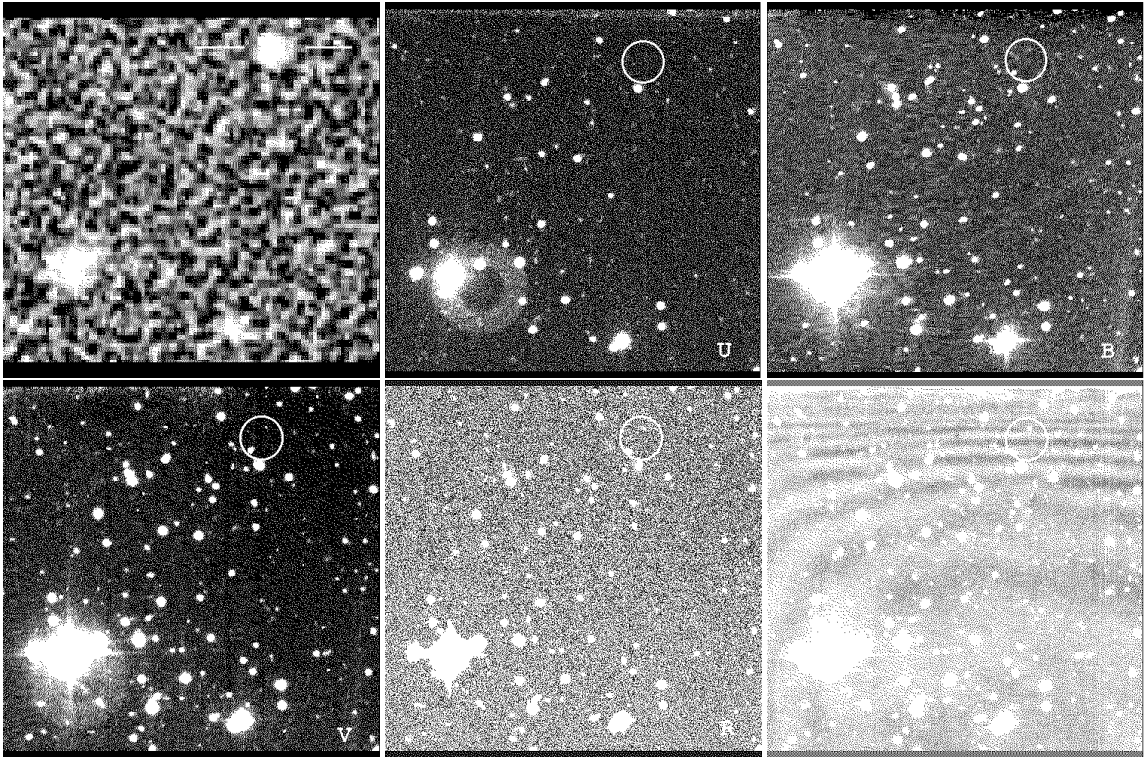

\begin{center}
  \resizebox{5cm}{5cm}{\includegraphics{7418.f8rp}}
  \resizebox{5cm}{5cm}{\includegraphics{7418.f9rp}}
  \resizebox{5cm}{5cm}{\includegraphics{7418.f10rp}}
  \resizebox{5cm}{5cm}{\includegraphics{7418.f11rp}}
  \resizebox{5cm}{5cm}{\includegraphics{7418.f12rp}}
  \resizebox{5cm}{5cm}{\includegraphics{7418.f13rp}}
    \caption{A set of images near the GRB 910122 field. The first image is a
      microscope image of plate AM~24589, showing the object (indicated
      between lines). The remaining 5 images were obtained with the Danish
      1.54-m Telescope at ESO La Silla Observatory in the U, B, V, R and I
      bands.  The circle represents the $1\sigma$ astrometrical error box
      at the object position. North is at the top and east to the left.}
\end{center}
\end{figure*}  

\begin{table}[h]
\begin{center}
\caption{Log of observations covering the GRB 910122 position with the 1.54-m Danish Telescope.}
  \begin{tabular}{lccccc}
    \hline Date & \multicolumn{5}{c}{Exposure time (s)} \\
    & U & B & V & R& I \\
    \hline 
    12 Sep 1996   & 1200 & -    & -   &  -  & -    \\
    13 Sep 1996   & -    & -    & 500 &  -  & 400  \\
    12 March 1997 & 1800 & 1200 & 900 & 600 & 600  \\
    15 March 1997 & -    & 1800 & 900 & 600 & 600  \\
    16 March 1997 & 6925 & 1800 & 900 & 600 & 600  \\
    26 June 1997  & -    & 3000 & -   & -   & -    \\
    27 June 1997  & 4800 & -    & -   & -   & -    \\
    28 June 1997  & 4800 & -    & -   & -   & -    \\
    29 June 1997  & 2400 & -    & 2700& 1200& 1200 \\
    4 July 1997   & 4500 & -    & -   & -   & -    \\
    5 July 1997   & -    & -    & -   & -   & 6000 \\
    \hline
    \multicolumn{6}{c}{Total:~
  15.59$^{hr}$~=~~26425$^{s}$+~7800$^{s}$~+~5900$^{s}$+~3000$^{s}$~+~9400$^{s}$}\\
    \hline
\end{tabular} 
\end{center}
\end{table}

\section{Discussion}
\subsection{Reliability of the object.}

There are different theories regarding the kind of silver grains structure
that OTs should leave in the emulsion.  Schaefer (1981) suggested that
flash images on large exposure plates are expected to be shallower than
those of stars, which can be attributed to the failure of the law of
reciprocity in photographic emulsions. Other authors claim that inadequate
guiding or variations in the atmospheric turbulence, could make the flash
images steeper (Moskalenko et al.  1989, Varady and Hudec 1992).  However
if the object image was produced by a long lasting ($\sim$ hours) optical
counterpart similar to OT/GRB 970228 and OT/GRB 970508, the structure of the
object would be indistinguishable from that of the stars on the plate.

When half of the plate AM 24589 was blinked, other two bright star-like
objects were found. However, taking into account the star-like faults rate
for AM series shown in Table~1, we would expect $\sim$10 star-like images,
which is comparable to the number of spots found in the blinking test.
Furthermore, the observing log does not contain any indication of a double
exposure, therefore a double exposure can be excluded with a high
probability.

Changing the focal length of the microscope, we were able to examine the
object at different deepness. In spite of its faintness, we realized that
the silver grains are geometrically extended in a cylinder shape up to the
emulsion/glass boundary, decreasing the density of grains for deeper
regions.  This is consistent with the structure of grains for real images
(Greiner et al.  1990a, \.{Z}ytkow 1990). We also detected a marginal
deposition of silver grains above the gelatine surface as it is expected
for exposed regions (Lau and Krug 1957).  Furthermore the inspection by
means of reflected light microscopy did not reveal distortions on any side
of the emulsion as it was detected in some plate faults (Greiner et al.
1990a).

The bright stars show coma distortion towards the centre of the plate,
however neither the object nor the faint stars showed any distortion (see
Fig~2).  In order to determine if the object shape was different than that
from the stars, the equidensity profiles of the object and the surrounding
stars were compared.  As it is shown in Fig 2, the object seems lightly
steeper than the stars. However, the profile of the object is very similar
to the faint stars, without coma distortion towards the plate centre.
Thus, we conclude that the study of the profiles does not show a
significant difference between the faint stars and the object.  {\em
  Therefore we have considered the star-like object as real, although the
  fault origin cannot be completely excluded.}

\subsection{The flash parameters}

For a given intensity $I$ and exposure time $T$, the photographic density
$D$ is expected to follow the Schwarzschild's reciprocity law
(Schwarzschild 1900), $D=I^{q}T$ , where $q$ is a function of the exposure
time $T$ (with $q \leq 1$ for a short flash and $q \geq1$ for a long one,
following Eastman Kodak, 1935).

If we assume that $q$ does not vary strongly with $T$, the li\-miting
magnitude $m_{flash}$ for a flash lasting $T_{flash}$ on a plate of
limiting magnitude $m_{plate}$ and exposure time $T_{plate}$ is:

$$m_{flash}=m_{plate}-(2.5/q)~log~(T_{plate}/T_{flash})$$

According to this formula, assuming the value of $q$ calculated by Schaefer
et al. (1981), and taking into account the magnitude of the object on the
plate (B~$ \sim$ 13.5 mag) and the exposure time of the plate (90 min), the
expected magnitude for a 1-s flash would be $\sim$ 6.0 mag. We note that OT
1959 (Moskalenko et al. 1989), OT 1966 (Greiner et al. 1991), OT 1901
(Schaefer et al. 1984), OT 1946 and OT 1961 (Hudec et al. 1990) reached
analogous magnitudes.

We still cannot rule out that the object on AM 24589 may have been caused
by a head-on meteor. If we assume a flash duration of $\sim$1 s, the
probability that the object found on the plate AM 24589 is due to a head-on
meteor can be roughly estimated;

Although the rate of meteors below 4 mag is not very well measured, it has
been  estimated that the frequency of $\sim$ 6 mag meteors is at
most $\sim $10 h$^{-1}$ sr$^{-1}$ (Karnashov et al.  1991). Thus, the
probability of detecting a $\sim$6 mag meteor inside the GRB 910122 IPN
error box (area $\sim$20 arcmin$^{2}$) on a plate like AM 24589 exposed 90
minutes, is $ < 3 \times 10^{-5}$.  However, as it is shown in the ``A''
plot of Fig~2, the object is not trailed with an upper limit of ~10 arc sec
trailing. So, the meteor should have approached within 10 arcsec off the
vertical.  Taking into account the estimation made by Schaefer et al.
(1981) for OT 1928 (an upper limit of 3 arc sec trailing), the probability
that a meteor will travel inside a cone with an aperture of ~10 arc sec
would be $\sim 10^{-7}$.

Therefore, the joint probability of both having on a AM plate a head-on
meteor and detecting the meteor in the IPN error box is at most $\sim 3
\times 10^{-12}$. Taking into account that the monitoring time of the
examined AM plates amounts to $\sim 4 \times 10^{3}$ hr , the total
probability of finding a $\sim$6.0 mag head-on meteor in the GRB 910122 IPN
error box in any AM plate used in this study, is below $ \sim 10^{-8}$.
Considering that a region $\sim$ 100 times larger than the 3$\sigma$ error
box was visually inspected, an upper limit to the probability of finding a
$\sim$6.0 mag head-on meteor around the GRB position is $\sim 10^{-6}$,
which is negligible.

We would like to point out that the former calculation has been performed
assuming the plate scale and the monitoring time of the AM series, which
represents a 84.5\% of the total monitoring time.  Other series would
impose lower upper limits to the probability because of their larger plate
scale in comparison to the AM series.

On the other hand, the possible presence of a minor planet has been checked
using data of the Minor Planet Center. As expected by the high ecliptic
latitude of the GRB ($\beta = -32^{\circ}$) there is no known minor planet
$15^{\prime}$ around the error box at the time of plate AM 24589 (Williams
1998).  The plate AM 24589 was taken in 1945, so any relationship with
satellite glints is obviously excluded. Astrometry of the object yielded:
$\alpha_{2000} =19^{h} 47^{m} 23^{s} \pm 2^{s}, \delta_{2000} = -70^{\circ}
33' 7.3'' \pm 9''$.

If the object is due to optical emission from the source that produced GRB
910122, and assuming a 1-s duration flash and the same gamma-ray fluence in
1945 and 1991, the corresponding gamma to optical (B~band) fluences ratio
is $S_{\gamma}/S_{opt} \sim 3 \times 10^{3}$, and the X-ray to optical (B
band) fluences ratio is $S_{\it x}/S_{opt} \sim 10^{2}$.

\subsection{A search for the quiescent optical counterpart of the object.}
We have tried to correlate the position of the object with catalogues of
sources (from X-rays to radio wavelengths), by means of the SIMBAD,
HEA\-SARC and NED databases.  The result is that no sources are found
within 30'' from the object position. Only the infrared source IRAS
19419-7038 is located 118'' away. Taking into account that the IRAS angular
resolution at $12~\mu m$ is 45" (Beichman et al. 1987), we conclude that
the relationship between this infrared source and the object is unreliable.

\begin{table*}[t]
\begin{center}
\caption{Magnitudes and positions for objects within the $3\sigma$
  astrometrical error circle of the possible OT found on the plate
  AM~24589.}
  \begin{tabular}{rllllllll}
 \hline

 Object& \multicolumn{2}{c}{Coordinates} & \multicolumn{5}{c}{Magnitudes} &
 Proposed \\ 

number& $ \alpha(2000)$&$ \delta(2000)$& U & B & V & R & I & class\\

 \hline                        

{\bf 1}&$19^{h}47^{m}22.36^{s}$&$-70^{\circ}33'0.9''$&$> 22.0 $&$23.9
\pm 0.5$&$22.30 \pm 0.20$&$21.15 \pm 0.15$&$20.00 \pm0.15$& K7/M0~{\rm V}\\
& & & & & & & & K5/M1~{\rm III} \\
{\bf 2}&$19^{h}47^{m}25.03^{s}$&$-70^{\circ}33'4.7''$&$> 22.0
$&$24.0\pm0.5$&$23.30\pm0.30$&$23.36\pm0.30$&$ > 22.00$&  \\
{\bf 3}&$19^{h}47^{m}18.56^{s}$&$-70^{\circ}33'3.9''$&$> 22.0
$&$23.4\pm0.4$&$22.89\pm0.25$&$21.35\pm0.30$&$20.20 \pm 0.25$&  \\
{\bf 4}&$19^{h}47^{m}24.35^{s}$&$-70^{\circ}33'5.7''$&$> 22.0 $&$24.0 
\pm0.5$&$23.05\pm0.25$&$22.01\pm0.10$&$21.50\pm0.30$&   \\
{\bf 5}&$19^{h}47^{m}26.91^{s}$&$-70^{\circ}33'4.6''$&$> 22.0 $&$>
24.2$&$ 23.50 \pm 0.50$&$23.30 \pm 0.40$&$ 21.70 \pm 0.40$&  \\
{\bf 6}&$19^{h}47^{m}27.42^{s}$&$-70^{\circ}33'11.8''$&$> 22.0 $&$22.5
\pm 0.3$&$21.40 \pm 0.20$&$21.20 \pm 0.15$&$20.80 \pm0.25$& \\
{\bf 7}&$19^{h}47^{m}26.79^{s}$&$-70^{\circ}33'14.9''$&$> 22.0 $&$22.9
\pm0.3$&$22.10\pm0.30$&$22.10\pm0.30$&$21.80 \pm 0.45$&  \\
{\bf 8}&$19^{h}47^{m}25.52^{s}$&$-70^{\circ}33'15.3''$&$> 22.0 $&$23.5
\pm0.5$&$23.10\pm0.25$&$21.50\pm0.30$&$20.60\pm0.25 $&  \\
{\bf 9}&$19^{h}47^{m}24.65^{s}$&$-70^{\circ}33'16.1''$&$> 22.0 $&$21.5
\pm0.2$&$20.00\pm0.20$&$19.04\pm0.15$&$18.60 \pm0.20 $&  K2/K4~{\rm III}\\
{\bf 10}&$19^{h}47^{m}23.47^{s}$&$-70^{\circ}33'26.7''$&$17.7
\pm0.2$&$17.9 \pm 0.2$&$17.05 \pm 0.15$&$16.50\pm0.15$&$16.20 \pm0.15$&
F5/F8~{\rm V}~?\\
 \hline      

\end{tabular} 
\end{center}
\end{table*}

\begin{figure}
  \resizebox{\hsize}{!}{\includegraphics{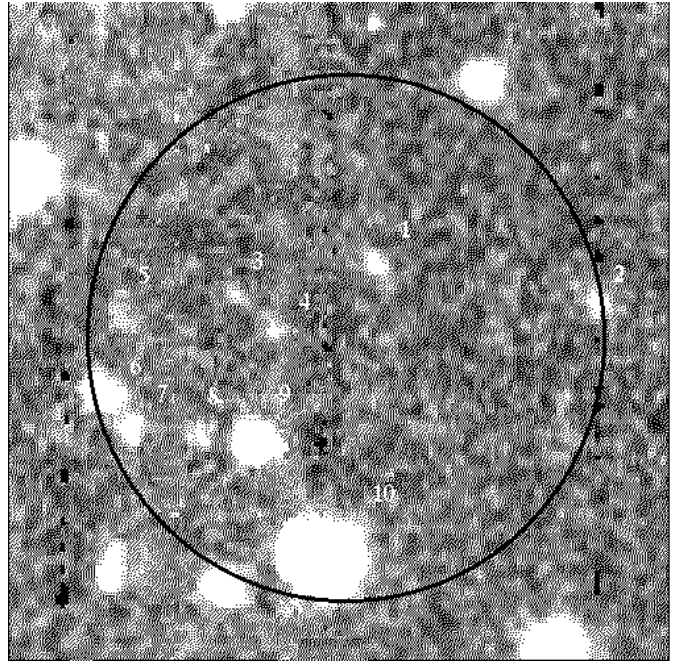}}
     \caption{The circle represents the $3\sigma$ astrometrical error circle 
       around the position where the object was found on plate AM~24589.
       Ten sources are consistent within the object error circle.  Only the
       objects \# 1, 3, and 4 are within the $1\sigma$ object error circle.
       The image is a combination of the U,B,V,R and I-band frames.
       Magnitudes and coordinates are given in Table 6.}
\end{figure}
UBVRI images of the field containing the OT are shown in Fig. 3. The
content of the $3\sigma$ astrometrical error circle is shown in Fig. 4.
Magnitudes and positions of all objects are displayed in Table 6.
Photometrical errors in the magnitudes are overestimated due to the
presence of several bad CCD columns.  Object \# 1 seems to be the most
interesting object within the 1$\sigma$ error circle, with U $> 22$, B = $
23.9 \pm 0.5$, V = $22.3 \pm 0.2$, R = $21.15 \pm 0.15$, I = $20 \pm 0.15$.
The colours are consistent with that of a K7/M0 dwarf star or a K5/M1
giant, although they could be also consistent with a reddened galaxy (see
Fig~5).  

The object on plate AM 24589 appears to be $\sim 0.6$ mag over the plate
limit and is absent on plate AM 24592 (obtained 3 hours and 5 minutes
after). Therefore, the object should have declined at a rate of $\geq 0.2$
mag/hr, which is similar to same decline rates in flares stars. Could
object \#1 be classified as a flare star?.
 
To estimate the probability of having a K7/M0 dwarf star or a K5/M1 giant
coincident with the 1$\sigma$ astrometrical error is difficult because of
various statistical biases in star catalogues.  Taking into account the
number of stars per square degree of magnitude B=22 at galactic latitude
$b^{\rm II}=-30^{\circ}$ and the fraction of K7/M0 spectral-type main
sequence and K5/M1 giant stars (Allen 1973), we expect $\sim 10^{-2}$ K7/M0
main sequence stars and $\sim10^{-4}-10^{-5}$ K5/M1 giants within a
$9^{\prime\prime}$ error circle (the size of the 1$\sigma$ astrometrical
error radius). Therefore, in the absence of stronger constraints we cannot
prove that this star has caused the object detected on the plate AM 24589.

Considering that in the solar neighbourhood the fraction of M0-M1 type
stars that can be considered as flare stars is ~5\% (Shakhovskaya 1994,
Mirzoyan 1994), the probability that a M0-M1 flare star is located by
chance inside the 1$\sigma$ astrometrical error would be $\lesssim
5\times10^{-4}$.  Therefore, if object \#1 was a flare star, it is likely
to be the responsible of the object found on the plate AM 24589.  In this
case, the flare amplitude would be $\Delta B \sim 10$ mag which is slightly
higher than the extreme flares of the CZ Cnc stars (Greiner and Motch 1994,
Toth et al.  1996).  It should be noted that M flare stars have been found
at OTs position inside two GRB error boxes (Greiner and Motch 1995, Hudec
et al.  1997). In any case an optical spectrum would be desirable in order
to clarify its nature.

\begin{figure}
  \resizebox{\hsize}{!}{\includegraphics[angle=-90]{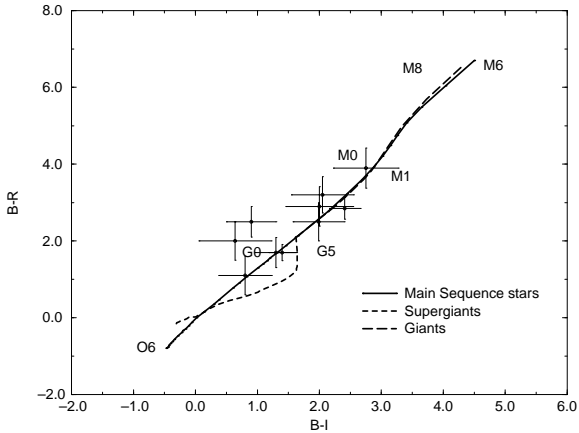}}
     \caption{Colour-colour diagram for the objects inside the  3$\sigma$
       error box shown in Fig. 4. The upper one is object \# 1 which is
       consistent with a K7/M0 star. Magnitudes are uncorrected for
       interstellar extinction.}
\end{figure}

\subsection{The lack of OTs within the GRB 910122 error box.}
OTs have been found at the edge or slightly outside GRB error boxes, like
GRB 910219 (Hudec et al. 1994b), GRB 790325b (Hudec et al.  1990) or GRB
791006b (Greiner et al. 1991). In fact several authors suggest a strong
relationship between quasars and OTs (Vrba et al. 1994, Hudec et al. 1997),
but statistical studies do not show a convincing correlation between
quasars and GRBs (Gorosabel et al. 1995, Schartel et al.  1997, Gorosabel
and Castro-Tirado 1998b, Gorosabel and Castro \-Tirado 1998c, Hurley et al.
1998).

The object is inside the WATCH error circle (Sazonov et al. 1998), but
$5^{\prime}$ outside the improved IPN error box for GRB 910122 (Hurley et
al. 1997) and not included in any of the three annuli determined by the
{\it Granat}, {\it PVO} and {\it Ulysses} spacecraft. Even if we assume an
extended halo radius equal to 150~kpc and a high-velocity neutron star (at
1000~km/s), its apparent motion on the sky would be only $\sim$ 0.14'' in
100 yr.  Therefore, theories based on high-speed neutron stars in the halo
will not allow to place the optical source that far away the IPN error box
(Li and Dermer 1992).  The maximum allowable distance to detect such proper
motion of $5^{\prime}$ in 46 yr (from 1945 to 1991) would be 35 pc, but
such a nearby neutron star would have been easily detected in previous
all-sky X-ray surveys.

Therefore we can set a lower limit of 0.55 yr for any recurrent optical
transient activity (above 14.1 mag) related to GRB 910122. This can be
compared with the 2.1 yr in the case of GRB 910219 (Hudec et al. 1994b).
Further results on the optical content of the GRB 910122 IPN error box will
be published elsewhere (Gorosabel, Castro-Tirado and Lund 1998).

\section{Conclusions}
We have examined $\sim$ 3995 plates at HCO for GRB 910122 covering $\sim$
0.55 yr.  No convincing evidence of optical transient emission was found
within the GRB 910122 error box. Additional plates at ROE did not show any
suspicious object either. However, a possible OT was found at HCO on the
plate AM 24589, located $5^{\prime}$ outside the improved error box
provided by the IPN.

Multicolour optical observations were carried out, finding 3 sources
consistent, within the $1\sigma$ object astrometrical error box. One of
them shows colours typical of a M dwarf star or a reddened galaxy. In any
case it seems to be unrelated to the GRB because of its location outside
the IPN error box.

\section*{Acknowledgments}
The authors warmly thank the Harvard College Observatory plate curator M.
L. Hazen and A. Doane for their help at the HCO plate stacks. We are
indebted to the referee R. Hudec for valuable comments and advices.  Also,
we would like to thank Prof.  Ofer Bar-Yosef, at the Stone Age Laboratory,
Department of Anthropology (Peabody Museum-Harvard University) to have
given us the opportunity of using the reflected-transmi\-tted light
microscope. The authors also wish to thank N. Lund for the facilities
given in order to perform the optical observations at La Silla, S. Tritton
for the support provided at The Royal Observatory of Edinburgh, and G.V.
Williams for the information provided by the IAU Minor Planet Center. We
are grateful to K.  Hurley for kindly providing the new coordinates of the
GRB 910122 error box, and to L.  Metcalfe, J. Greiner and B.  Montesinos
for fruitful conversations.  This work has been partially supported by
Spanish CICYT grant ESP95-0389-C02-02.

\end{document}